\documentclass[pra,aps,twocolumn,showpacs,superscriptaddress]{revtex4-1}

\usepackage{graphicx}
\usepackage{bm,color}

\usepackage{graphicx}
\usepackage{bm,color}
\usepackage{multirow}

\newcommand{\be}{\begin{eqnarray}}
\newcommand{\ee}{\end{eqnarray}}

\renewcommand{\theequation}{\arabic{equation}}
\begin{document}

\title{Phase structure of the interacting Su-Schrieffer-Heeger model and the relationship with the Gross-Neveu model on lattice}
\date{\today}

\author{Yoshihito Kuno}
\affiliation{Department of Physics, Graduate School of Science, Kyoto University, Kyoto 606-8502, Japan}

\begin{abstract}
The $N$-flavor interacting Su-Schrieffer-Heeger (i-SSH) model realizable in cold-atoms in an optical lattice is studied.
We clarify the relationship between the i-SSH model and the Chiral-Gross-Neveu-Wilson (CGNW) model. 
Following the previous study of the CGNW model in the high-energy physics community, 
the groundstate phases of the i-SSH model are investigated and interpreted from the view of the phases of the CGNW model.  
The interaction effect on the i-SSH model, belonging to the topological BDI class, is grasped by following the view of the dynamical breakdown of chiral symmetry in the CGNW model. 
Furthermore, we compare the large-$N$ groundstate phase diagram with that of the $N=1$ case obtained by exact diagonalization 
and then propose a table-top cold-atom quantum simulator to test the model. 
\end{abstract}


\maketitle
\section{Introduction}
The topological condensed matter model is deeply related to the high-energy physics model in a lattice. 
In particular, topological insulators are known to be related to Dirac fermions in a lattice \cite{Shen,Fradkin}, 
which is a major component in high-energy physics in a lattice \cite{Wilson,Rothe,Creutz2}. 
Investigation of the relationship between topological condensed matter model and high-energy physics model on lattice 
leads to deep understanding of the phases of matter in the topological condensed matter model. 
With the help of high-energy physics study, 
there is a new possibility to understand the strongly correlated topological model and its novel phase structure.
Such an interdisciplinary research can give us important insights into strongly correlated topological systems. 
For example, recently, a relationship between a cold-atom condensed matter model with a non-trivial topological phase 
and a high-energy physics model has been discussed \cite{Bermudez,Cirac,Zache,Kuno}. 
Such an approach also gives us deep understanding of the topological condensed matter model, which is realizable in cold-atom systems.  
However, interdisciplinary study of strongly correlated topological systems is still lacking. 
Thus, in this work, motivated by previous studies \cite{Bermudez,Aoki,Araki}, 
we study a fundamental topological model with interactions, the interacting Su-Schrieffer-Heeger (i-SSH) model \cite{SSH,Asboth},
and show that the i-SSH model has a clear relationship with the Chiral-Gross-Neveu-Wilson (CGNW) model, 
which has been extensively studied in the high-energy physics community \cite{Gross-Neveu,Aoki,Creutz} because the model 
has common features of lattice quantum chromodynamics (QCD) \cite{Rothe}. 
In high-energy physics, the $N$-flavor CGNW model has been analyzed using the large-$N$ expansion and turned out to possess a rich phase diagram \cite{Aoki}.
Following the study, the $N$-flavor (component) i-SSH model is studied using the large-$N$ expansion. 
In particular, we study how topological phases are affected by interaction. 
The i-SSH model exhibits a rich phase diagram induced by interactions. 
The groundstate phase diagram has clear correspondence to that of the CGNW model. 
Furthermore, we investigate the $N$-flavor dependence of the model, 
and then propose implementation schemes to realize the i-SSH model in cold-atoms in an optical lattice.

The paper is organized as follows. In Sec. II, our target models is introduced. 
In Sec. III, we show the relationship between the i-SSH model and the CGNW model.  
In Sec. IV, we explain the large-$N$ calculation and show the large-$N$ groundstate phase diagram of the i-SSH model. 
In Sec. V, we carry out an exact diagonalization for the i-SSH model and obtain global phase diagrams of the single flavor case of the i-SSH model, and then compare the result to the large-$N$ result. In Sec.VI,  we discuss the implementation scheme of the i-SSH model by using recent cold-atom experimental techniques. Finally, the conclusion is given in Sec. VII.

\section{$N$-flavor SSH model and CGNW model}
We start with the $N$-flavor Su-Schrieffer-Heeger (SSH) model \cite{SSH,Asboth},
\begin{eqnarray}
&&H^{N}_{S}=-\sum_{i}\sum^{N}_{\alpha=1}(J_{1}a^{\dagger}_{\alpha,i}b_{\alpha,i}+J_{2}a^{\dagger}_{\alpha, i+1}b_{\alpha,i}+\mbox{h.c.}),
\label{SSHN}
\end{eqnarray}
where $a^{(\dagger)}_{\alpha,i}$ and $b^{(\dagger)}_{\alpha,i}$ are annihilation (creation) operators for the left and right inner site in a unit cell $i$, $\alpha$ is the flavor index, and
$J_{1(2)}$ is the inner (inter) site hopping amplitude.
In this work, we consider two types of SU($N$) symmetric interaction $V_{{\rm I (II)}}$,
\begin{eqnarray}
&&V_{{\rm I}}=-\frac{U}{2N}\sum_{i}\biggr[\sum^{N}_{\alpha=1}(n^{a}_{\alpha,i}-n^{b}_{\alpha,i})\biggl]^{2},\nonumber\\
&&V_{{\rm II}}=V_{{\rm I}}-\frac{U}{2N}\sum_{i}\biggr[\sum^{N}_{\alpha=1}(n^{a}_{\alpha,i+1}-n^{b}_{\alpha,i})\biggl]^{2},\nonumber
\label{SSHint}
\end{eqnarray}
where $n^{a(b)}_{\alpha,i}=a^{\dagger}_{\alpha,i}a_{\alpha, i} (b^{\dagger}_{\alpha,i}b_{\alpha, i})$ is the particle number operator and
$U$ is the interaction strength. 
The above interactions may be realized in a cold-atom experimental system \cite{Cazalilla2,Taie,Gorshkov}. 
For the $N>1$ case, though attractive on-site interactions between different flavors and repulsive nearest-neighbor (NN) interactions appear, 
there is a possibility to tune these interactions by combining recent experimental techniques, 
e.g., Feshbach and orbital-Feshbach resonance \cite{Inouye,Hofer}, and dipole-dipole interaction (DDI) \cite{Ferlaino, interaction}. 
For the case $N=1$ (single component case), the situation is quite simple. $V_{\rm I}$ reduces to a repulsive interaction between NN sites in the same unit cell $i$ 
and $V_{\rm II}$ reduces to a repulsive interaction appearing in all pairs of NN sites. 
Here, the i-SSH model is defined as $H^{N}_{S}+V_{{\rm I (II)}}$.  
In what follows, we call the Hamiltonian $H^{N}_{S}+V_{{\rm I}({\rm II})}$ the type-I (II) i-SSH model . 
In the context of condensed system physics, the type-II interaction is related to the  z-component Hund's rule coupling in spin $N/2$ system \cite{Affleck}.

The bulk-momentum Hamiltonian of Eq.~(\ref{SSHN}) for a certain flavor $\alpha$ is given by
$h^{S}_{\alpha}(k)=[-J_{1}-J_{2}\cos k]\hat{\sigma}_{x}+[-J_{2}\sin k]\hat{\sigma}_{y}$.
Then, using a spinor field $f_{\alpha}(k)=(a_{\alpha}(k),b_{\alpha}(k))^{t}$, the second quantization form is written as $\sum^{N}_{\alpha=1}\int \frac{dk}{2\pi}f^{\dagger}_{\alpha}(k)h^{S}_{\alpha}(k)f_{\alpha}(k)$. This form is used in the large-$N$ expansion.

Next, we consider the $N$-flavor CGNW model \cite{Bermudez,Aoki,Gross-Neveu}. The model is written using Wilson fermions \cite{Wilson,Rothe}. Then, the model includes an additional NN hopping term, called the Wilson term, parametrized by $r$, called the Wilson parameter \cite{Wilson,Rothe,Bermudez,Aoki}. 
The model is given by 
\begin{eqnarray}
&&H^{N}_{G}=\sum_{i}\sum^{N}_{\alpha=1}\biggl[\frac{1}{2}(\bar{\psi}_{\alpha,i}(-i\gamma^{1})\psi_{\alpha,i+1}+\mbox{h.c.})\nonumber\\
&&+m_{0}\bar{\psi}_{\alpha,i}\psi_{\alpha,i}-\frac{r}{2}(\bar{\psi}_{\alpha}\psi_{\alpha,i+1}+\mbox{h.c.})\biggl]\nonumber\\
&&-\frac{g^{2}}{4N}\sum_{i}\biggl[\biggl(\sum^{N}_{\alpha=1}\bar{\psi}_{\alpha,i}\psi_{\alpha,i}\biggl)^{2}
-\biggr(\sum^{N}_{\alpha=1}\bar{\psi}_{\alpha,i}\gamma^{5}\psi_{\alpha,i}\biggl)^{2}\biggr],
\label{CGNW}
\end{eqnarray}
where $\psi_{\alpha,i}$ is the spinor field with a flavor $\alpha$ on lattice site $i$, and the gamma matrices are set as $\gamma^{0}=\hat{\sigma}_{z}$, 
$\gamma^{1}=-i\hat{\sigma}_{y}$, $\gamma^{5}=\hat{\sigma}_{x}$, 
and $\bar{\psi}_{\alpha,i}=\psi^{\dagger}_{\alpha,i}\gamma^{0}$. $m_{0}$ is the effective mass, defined as $m_{0}\equiv m+\frac{r}{2}$, where $m$ is the Wilson mass. 
$g^{2}$ is the coupling constant of the interaction that is invariant for continuous chiral symmetry transformation \cite{CChiral}.
In this study, we set the lattice spacing to unity and set $r=1$. 
Then, the bulk-momentum Hamiltonian of the non-interacting part of $H^{N}_{G}$ for a flavor $\alpha$ is given by $h^{G}_{\alpha}(k)=[m+1-\cos k]\hat{\sigma}_{x}+[\sin k]\hat{\sigma}_{y}$. 
The dispersion of $h^{G}_{\alpha}(k)$ with $r\neq 0$ avoids having zero energy at $k=\pm \pi$; thus, the fermion doubler is eliminated \cite{Wilson,Rothe}.

\section{Relationship}
There is a clear relationship between the type-I i-SSH model and the CGNW model. 
The left and right inner site in a unit cell in the type-I i-SSH model correspond to the color degrees of freedom of the Wilson fermion in the CGNW model. 
There exists a clear correspondence between the gamma matrices in $h^{G}_{\alpha}(k)$ 
and the Pauli matrices in $h^{S}_{\alpha}(k)$: 
$\gamma^{0}\longleftrightarrow \hat{\sigma}_{x}$, $\gamma^{1}\longleftrightarrow -i\hat{\sigma}_{z}$, 
and $\gamma^{5}\longleftrightarrow \hat{\sigma}_{y}$.
Furthermore, by imitating the form of the interaction in Eq.~(\ref{CGNW}) 
we can deform $V_{\rm I}$ in the type-I i-SSH model into
\begin{eqnarray}
V_{\rm I}= -\frac{U}{4N} \sum_{i}\biggr[\biggr(\sum^{N}_{\alpha=1}f^{\dagger}_{\alpha,i}\hat{\sigma}_{x}f_{\alpha,i}\biggl)^{2}
-\biggr(\sum^{N}_{\alpha=1}f^{\dagger}_{\alpha,i}i\hat{\sigma}_{z}f_{\alpha,i}\biggl)^{2}\biggl], \nonumber\\
\label{SSHint}
\end{eqnarray}
where $f_{\alpha,i}$ is a spinor field $f_{\alpha,i}=(a_{\alpha,i},b_{\alpha,i})^{t}$.  
By comparing Eq.(\ref{SSHint}) with the form of the interaction in Eq.(\ref{CGNW}), there are operator relations between the type-I i-SSH model and the CGNW model: 
\begin{eqnarray}
&&f^{\dagger}_{\alpha,i}\hat{\sigma}_{x}f_{\alpha,i} \longleftrightarrow \bar{\psi}_{\alpha,i}\psi_{\alpha,i}, \label{Chiralrelation}\\
&&f^{\dagger}_{\alpha,i}i\hat{\sigma}_{z}f_{\alpha,i}\longleftrightarrow \bar{\psi}_{\alpha,i}\gamma^{5}\psi_{\alpha,i}. \label{Aokirelation}
\end{eqnarray}
These relations indicate that the inner-bond operator in the i-SSH model corresponds to the particle-anti-particle pairing operator in the CGNW model and 
the density-difference operator between the left and right inner site in a unit cell to the pseudo-scalar operator, which corresponds to a pion field and whose expectation value characterizes a pion condensation in the high-energy physics context \cite{Rothe,Aoki,Gross-Neveu}.

In high-energy physics study, the CGNW model with $m=0$ and $r=0$ has been expected to have non-zero expectation value of $\bar{\psi}_{\alpha,i}\psi_{\alpha,i}$ 
due to the interaction $g^{2}$, which is known as the spontaneous dynamic breakdown of chiral symmetry \cite{Gross-Neveu,Creutz2}. 
Here, because our CGNW model is assumed to have a finite mass $m_{0}\neq 0$, the model does not exhibit such a spontaneous chiral symmetry breaking. 
However, the dynamical effect induced by the interaction $g^{2}$ affects the value of the mass term $m_{0}$. 
Then, from the relation of Eq.~(\ref{Chiralrelation}) and by comparing $h^{G}_{\alpha}(k)$ with $h^{S}_{\alpha}(k)$, 
we expect that in the i-SSH model, the same mechanism leads to a modification of the parameter $J_{1}$, 
that is, interaction acts as correction for the parameter $J_{1}$, which determines the strength of the inner-bond order in the i-SSH model.

Some previous studies \cite{Aoki,Creutz} have expected that 
the CGNW model has a novel state with a non-zero expectation value of $\bar{\psi}_{\alpha,i}\gamma^{5}\psi_{\alpha,i}$ in a large $g^{2}$ regime. 
This state is known as the Aoki phase, which is a parity-broken phase \cite{Creutz2,Aoki,Creutz}.
Then, from the relation of Eq.~(\ref{Aokirelation}), we expect that the Aoki phase corresponds to the density-wave phase in the i-SSH model.
In what follows, from a unified perspective, we also call the density-wave order in the i-SSH model the Aoki phase.

The type-II i-SSH model can be also related to the lattice version of an extended Gross-Neveu model, which has non-local interactions and has been discussed in a high-energy physics context \cite{Fuertes,Reisz}. The $V_{\rm II}$ term can also be deformed in the same way as Eq.~(\ref{SSHint}). 
The details are explained in the Supplemental Material \cite{Sup}. 

\section{Large-$N$ expansion}
\begin{figure}[t]
\begin{center} 
\includegraphics[width=8cm]{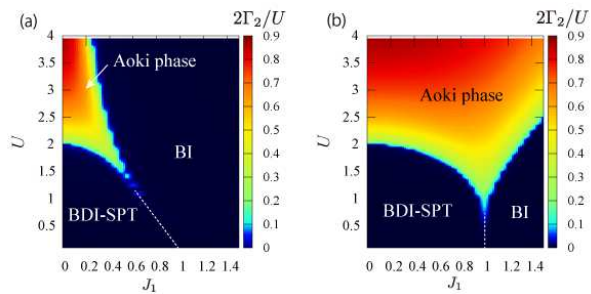}
\end{center} 
\caption{Large-$N$ phase diagram: (a) Type-I i-SSH model,
(b) Type-II i-SSH model. For both cases, $J_{2}=1$ and three phases appear: the BI phase, the BDI-SPT phase (for odd $N$ case), and the Aoki phase.}
\label{PD}
\end{figure}
\begin{table}[b]
\caption{ Phase correspondence between the i-SSH model and the CGNW model.}
\begin{tabular}{|c|c|}
\hline
i-SSH model & CGNW model \\ \hline
\begin{tabular}[c]{@{}c@{}}Band-insulator phase\\ (Inner-bond order)\end{tabular} & \multirow{2}{*}{\begin{tabular}[c]{@{}c@{}}Chirally-broken phase\\ (Particle-anti-particle\\ pair condensation)\end{tabular}} \\ \cline{1-1}
\begin{tabular}[c]{@{}c@{}}BDI-SPT phase\\ (Inter-bond order)\end{tabular} &  \\ \hline
\begin{tabular}[c]{@{}c@{}}Density wave phase\\ \end{tabular} & \begin{tabular}[c]{@{}c@{}}Parity-broken Aoki phase\\ (Pseudoscalar condensation)\end{tabular} \\ \hline
\end{tabular}
\label{correspondence}
\end{table}
The large-$N$ expansion has succeeded capturing the groundstate phase diagram of the CGNW model in high-energy physics \cite{Gross-Neveu,Aoki,Creutz}. 
Motivated by this fact, we apply the large-$N$ expansion to both type-I and II i-SSH model. 
According to the classification of the non-interacting topological Hamiltonian \cite{Schnyder,Ryu,Kitaev}, the SSH model is classified in the BDI class. 
The Hamiltonian $h^{S}_{\alpha}(k)$ has chiral ($S$), time-reversal ($T$), and charge-conjugation symmetry ($C$) \cite{Chiral}.  
In addition, if an odd $N$-flavor SSH model is assumed, the model possesses a symmetry-protected topological (SPT) phase \cite{Senthil, Bermudez,Sirker}. 

We investigate how interactions change the topological phase structure of the i-SSH model and break the BDI symmetry. 
Let us focus on the application of the large-$N$ expansion to the type-I i-SSH model (The detailed treatment is given in the Supplemental Material \cite{Sup}).
In the large-$N$ expansion, the $V_{\rm I}$ term in the type-I i-SSH model can be decoupled by introducing an auxiliary mean-fields $\Gamma_{1(2)}$. These mean fields are introduced in employing the Hubbard-Stratnovich transformation for $V_{\rm I}$ in the process of the large-$N$ calculation when assuming the translational symmetry of the system. $\Gamma_{1}$ and $\Gamma_{2}$ correspond to the expectation values $\langle f^{\dagger}_{\alpha,i}\hat{\sigma}_{x}f_{\alpha,i}\rangle$ and 
$\langle f^{\dagger}_{\alpha,i}i\hat{\sigma}_{z}f_{\alpha,i}\rangle$, respectively.
Then, $\Gamma_{1}$ and $\Gamma_{2}$ can be incorporated into the Hamiltonian $h^{S}_{\alpha}(k)$. 
The effective bulk-momentum Hamiltonian is given by 
$h^{\rm I}_{\alpha}(k)=[-(J_{1}+\Gamma_{1})-J_{2}\cos k]\hat{\sigma}_{x}+[-J_{2}\sin k]\hat{\sigma}_{y}+\Gamma_{2}\hat{\sigma}_{z}$.
Practically, the value of $\Gamma_{1(2)}$ is determined by solving a saddle point equation parameterized by $J_{1}/J_{2}$ and $U/J_{2}$ \cite{Sup}. 
Here, it is clear that $\Gamma_{1}$ modifies the coupling $J_{1}$ as $\tilde{J}_{1}=J_{1}+\Gamma_{1}$. 
If $\Gamma_{1}>0$, $\Gamma_{1}$ acts as an enhancing effect for $J_{1}$. 
Conversely, $\Gamma_{2}$ contributes to the breakdown of BDI symmetry and leads to the Aoki phase. 
For the Hamiltonian $h^{\rm I}_{\alpha}(k)$, if $\Gamma_{2}\neq 0$, 
$h^{\rm I}_{\alpha}(k)$ is no longer in the BDI class because the $\hat{\sigma}_{z}$ term in $h^{\rm I}_{\alpha}(k)$ breaks $S$ symmetry. 
Therefore, if there exists a mean-field solution with $\Gamma_{2}\neq 0$, the type-I i-SSH model is not BDI class.  
This leads the system to not possess a non-trivial topological phase simultaneously with the appearance of the Aoki phase.

By solving numerically the saddle point equation derived from the large-$N$ expansion, 
we obtain the groundstate phase diagrams for both type-I and type-II i-SSH models, as shown in Fig.~\ref{PD}. 
For both cases, three phases appear: the band-insulator (BI), the BDI-SPT phase, and the Aoki phase. 
Here, the phase boundary between the BI and the BDI-SPT phase 
is determined by sgn($\tilde{J}_{1}-J_{2}$), 
i.e., if $\tilde{J}_{1}>J_{2}$ ($\tilde{J}_{1}<J_{2}$), the BI (the BDI-SPT) phase appears. 
The Aoki phase is characterized by $|2\Gamma_{2}/U|> 0$.
The type-I i-SSH phase structure in Fig.~\ref{PD} (a) perfectly corresponds to the phase structure of the previous study for the CGNW model \cite{Bermudez}.
The BDI-SPT phase is robust up to some extent of interaction strength $U$.  
Furthermore, through the value of $\Gamma_{1}$, the $V_{\rm I}$ acts as an enhancing effect for $J_{1}$, i.e., the inner-bond order (the BI phase) is enhanced. 
This appears in the result in Fig.~\ref{PD} (a): 
The phase boundary-line between the BI and the BDI-SPT phase in Fig.~\ref{PD} (a) is not on the line $J_{1}=J_{2}$ with increasing $U$, but left-tiled. 
For the weak $J_{1}$ regime, the BDI-SPT phase directly transitions to the Aoki phase with increasing $U$ because the Aoki phase is energetically favorable compared with creating the BI phase.   
Conversely, for the type-II i-SSH model, Figure.~\ref{PD} (b) indicates that the enlargement of the Aoki phase compared with the type-I results in Fig.~\ref{PD} (a) and that there is a direct phase transition from the BI to the Aoki phase with increasing $U$. Also, the BDI-SPT phase is robust up to $U/J_{2}\sim 3$. 
Although the $V_{\rm II}$ acts as a correction effect for both $J_{1}$ and $J_{2}$ as in the type-I interaction $V_{\rm I}$, 
this does not change the phase boundary-line $J_{1}=J_{2}$ between the BI and the BDI-SPT phase.

The correspondence of the phases between the i-SSH model and the CGNW model is summarized in table \ref{correspondence}. 
Next, we investigate the $N=1$ case to compare with the large-$N$ result obtained here.

\begin{figure}[t]
\begin{center} 
\includegraphics[width=6.5cm]{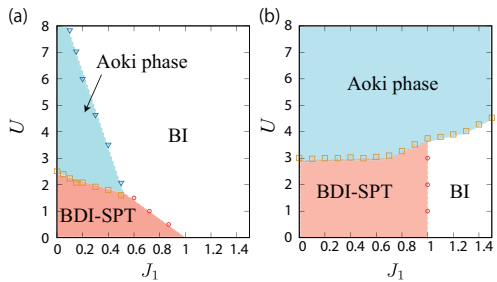}
\end{center} 
\caption{ $N=1$ phase structures obtained by exact diagonalization: 
(a) The type-I i-SSH model and (b) the type-II i-SSH model. For both cases, $J_{2}=1$.} 
\label{PDN=1}
\end{figure}

\section{$N=1$ groundstate phase diagram}
Using exact diagonalization, we investigate the groundstate phase diagrams of the type-I and type-II i-SSH models with $N=1$, 
where the number of lattice sites is $L=12$, $16$, and $20$ with periodic boundary conditions at half-filling, and we employed the Lancozs algorithm \cite{EDtext1,EDtext2} and finite size scaling. 
The obtained phase structures are shown in Fig.~\ref{PDN=1}. 
Compared with Fig.\ref{PD}(a), in Fig~\ref{PDN=1} (a) the phase boundary between the BDI-SPT phase and the Aoki phase rises for the small $J_{1}$ regime. The same behavior has been reported in the CGNW model \cite{Bermudez}. In particular, our numerics indicate the rise at $J_{1}=0$ is smaller than that of the CGNW model case \cite{Bermudez}.
For Fig.~\ref{PDN=1} (b), the phase boundary of the Aoki phase is lifted as a whole compared with Fig.\ref{PD}(b).
In particular, the tricritical point is lifted compared with Fig.~\ref{PD} (b). 
We expect that this may be caused by quantum fluctuation effects. However, the details will be studied in future work. The tricritical point in Fig.~\ref{PDN=1} (b) is in agreement with a previous study \cite{Sirker}.
After all, we conclude that the $N=1$ results for the type-I and type-II i-SSH model have qualitative agreement with the large-$N$ results in Fig.\ref{PD}. 
In addition, for Fig.~\ref{PD} (a) and (b), the critical behavior toward the Aoki phase is estimated by calculating the order parameter of the Aoki phase $O_{DW}$ and using finite-size scaling \cite{scaling}.
Our numerical calculation indicates that the universality class belongs to the $d=2$ Ising type, and the critical exponents of $O_{DW}$ take $\beta=1/8$ and $\nu=1$; 
the critical behavior in both type-I and type-II i-SSH models corresponds to the result of the phase transition between the Aoki phase and the BDI-SPT phase 
in the CGNW model \cite{Bermudez}. The details are shown in the Supplemental Material \cite{Sup}.  

\begin{figure}[t]
\begin{center} 
\includegraphics[width=9.5cm]{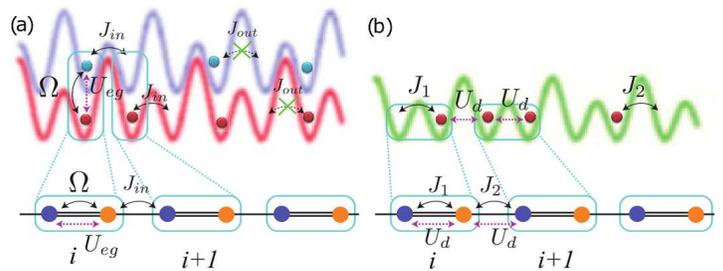}
\end{center} 
\caption{Implementation scheme using cold-atoms in an optical lattice: (a) the type-I i-SSH model, (b) the type-II i-SSH model.
In the type-I case, the interaction $V_{\rm I}$ appears as on-site s-wave scattering interaction $U_{eg}$ 
between the different internal states of fermionic atoms.
In the type-II case, the interaction $V_{\rm II}$ is implemented as long-range DDI $U_{d}$ using a dipolar fermionic atom.}
\label{Setup}
\end{figure}

\section{Implementation scheme for cold-atom experiments}
There are two types of implementation scheme for the type-I and type-II i-SSH model. 
In this section, we argue the implementation for the single flavor case $N=1$. 
Actually, a recent cold-atom experiment realized the standard SSH model (non-interacting) model by using optical super lattice setup \cite{Atala}, 
and the SSH model defined on a momentum-space lattice was realized in a cold-atom experiment \cite{Gadway}. 
Also, Ref.\cite{Song} reported the realization of another topological model related to the i-SSH model on a spin dependent one-dimensional optical lattice.
 
To realize the type-I i-SSH model in experiments, we employ two-different internal states of fermionic atoms and prepare two kinds of double-well optical lattice, shown as the blue and green colored lattice potentials in Fig.\ref{Setup} (a).  Each double-well optical lattice is fixed on the same one-dimensional spatial axis. Each double-well optical lattice is misaligned by one site with respect to each other, as shown in Fig.~\ref{Setup} (a). This system can be feasible using a spin-dependent optical lattice technique \cite{SPL1}. Here, each fermion can be independently trapped for each double-well optical lattice. For this lattice geometry, we add the Rabi coupling $\Omega$ by adding an external laser light. The Rabi coupling exchanges the two different internal states of fermions on same place \cite{synthetic}. 
The $\Omega$ can be regarded as the hopping $J_{1}$ in the SSH model. Then, we set a deep double-well situation for both optical lattices. 
This situation suppresses the hopping between NN unit cells denoted by $J_{out}$ in Fig.\ref{Setup} (a). The system only remains the hopping in a double-well, denoted by $J_{in}$ in Fig.~\ref{Setup} (a).  
Then, $J_{in}$ can be regarded as $J_{2}$ in the SSH model. 
Furthermore, in this system, an on-site interaction between the two different internal states of atoms denoted by $U_{eg}$ can be implemented because the two different internal states of fermionic atoms are spatially trapped at the same position. 
$U_{eg}$ can be regarded as $U$ in the type-I SSH model. Thus, the $V_{\rm I}$ term is realized and we obtain the type-I i-SSH model in this system. 
Because the type-I i-SSH model is directly connected to the CGNW model, 
the table top experimental simulator of the type-I SSH model has the possibility to become a quantum simulator of the CGNW model.

Conversely, to realize the type-II i-SSH model, a single fermionic atom with a large magnetic dipole moment is suitable. 
We prepare a one-dimensional double-well optical lattice to trap the atoms. The schematic figure is shown in Fig.\ref{Setup} (b). 
Here, the lattice geometry directly generates $J_{1}$ and $J_{2}$ hopping terms in the SSH model. 
Then, the large magnetic dipole moment of the atom can generate the DDI between NN sites denoted by $U_{d}$ in Fig.~\ref{Setup} (b), 
corresponding to $U$ in $V_{\rm II}$ if all dipole moments are polarized using external magnetic fields. 
In real experiments, ${}^{167}$Er \cite{Ferlaino} and ${}^{161}$Dy \cite{Lev} degenerate fermi gasses 
are candidates to realize the above setup because they have large magnetic dipole moments. 
A concrete parameter estimation for the two implementation schemes is given in the Supplemental Material \cite{Sup}. 
Our proposed experimental setups cover our target parameter regime for $J_{1}/J_{2}$ and $U/J_{2}$, as shown in Fig.\ref{PD} and \ref{PDN=1}. 

\section{Conclusion}
We studied an $N$-flavor i-SSH model and clarified the relationship with the CGNW model. 
For the i-SSH model, the large-$N$ expansion was carried out. We shown how interaction changes the phase boundary of the BDI-SPT phase and the Aoki phase. 
The interaction effect appears as a correction for the hopping amplitudes in the SSH model. 
This mechanism is analogous to the dynamical breakdown of chiral symmetry in the Gross-Neveu model.
Furthermore, interactions lead to the breakdown of the $S$ symmetry in the i-SSH Hamiltonian. This makes the i-SSH model out of the BDI class at a certain threshold value U and leads to the Aoki phase. This indicates that the $S$ symmetry breaking is related to the appearance of the Aoki phase. 
The phase diagram of the i-SSH model with $N=1$ was also calculated and was compared with the large-$N$ result. 
The phase diagrams show qualitative agreement with the large-$N$ result.  
Furthermore, we proposed an implementation scheme to realize the i-SSH model in future experiments.

\section*{Acknowledgments}
Y. K. acknowledges the support of a Grant-in-Aid for JSPS
Fellows (No.17J00486).

\clearpage  
\renewcommand{\theequation}{S\arabic{equation}}
\renewcommand{\thefigure}{S\arabic{figure}}
\renewcommand{\bibnumfmt}[1]{[S#1]}
\renewcommand{\citenumfont}[1]{S#1}
\setcounter{equation}{0}
\setcounter{figure}{0}
\section*{Supplemental Material}

\section*{A. Large-$N$ expansion}
\subsection*{Procedure}
We explain the large-$N$ expansion in detail. 
In particular, we show the procedure for the type-II i-SSH model, which can be directly reduced to the treatment for the type-I i-SSH model. 
We introduce the continuous imaginary-time $\tau$ and Grassman fields $A^{(*)}_{\alpha,i}$ and $B^{(*)}_{\alpha,i}$ for the operators $a^{(\dagger)}_{\alpha,i}$ and $b^{(\dagger)}_{\alpha,i}$, respectively. 
Then, by introducing a spinor field $\Psi_{\alpha,i}=(A_{\alpha,i},B_{\alpha,i})^{t}$, the partition function for the type-II i-SSH model $Z$ can be written as 
\begin{eqnarray}
Z&=&\int[dA^{*}_{\alpha,i}][dA_{\alpha,i} ][dB^{*}_{\alpha,i}][dB_{\alpha,i} ]  e^{-S},\nonumber\\
S&=&\int^{\beta}_{0}d\tau \biggr[\sum_{\alpha,i}\Psi^{*}_{\alpha,i}\partial_{\tau}\Psi_{\alpha,i}+ H^{N}_{S}(A^{*}_{\alpha,i},A_{\alpha,i},B^{*}_{\alpha,i},B_{\alpha,i})\nonumber\\
&&+V_{\rm II}(A^{*}_{\alpha,i},A_{\alpha,i},B^{*}_{\alpha,i},B_{\alpha,i})\biggl],
\end{eqnarray}
where $\beta$ is the inverse temperature. Here, the $V_{\rm II}$ term can be written in the following form,
\begin{eqnarray}
V_{\rm II}&=&-\frac{U}{4N} \sum_{i}\biggr[\biggl(\sum^{N}_{\alpha=1}\Psi^{*}_{\alpha,i}\hat{\sigma}_{x}\Psi_{\alpha,i}\biggl)^{2}
+\biggr(\sum^{N}_{\alpha=1}\Psi^{*}_{\alpha,i}\hat{\sigma}_{z}\Psi_{\alpha,i}\biggl)^{2} \nonumber\\
&+&\biggr(\sum^{N}_{\alpha=1}K^{*}_{\alpha,i}\hat{\sigma}_{x}K_{\alpha,i}\biggl)^{2}+\biggr(\sum^{N}_{\alpha=1}K^{*}_{\alpha,i}\hat{\sigma}_{z} K_{\alpha,i}\biggl)^{2}\biggl],
\end{eqnarray}
where $K_{\alpha,i}$ is a spinor field $K_{\alpha,i}=(A_{\alpha,i+1},B_{\alpha,i})^{t}$.

Using the Hubbard-Stratonovich transformation (HST), the $V_{\rm II}$ sector in the partition function $Z$ can be written as
\begin{eqnarray}
e^{-V_{\rm II}}&=&\int \prod^{4}_{\ell=1}[d\Gamma_{\ell,i}(\tau)]e^{-{V}_{\rm eff}(\Gamma_{\ell,i})},\nonumber\\
V_{\rm eff}&=&\sum_{i}\sum^{N}_{\alpha=1}\biggr[\frac{1}{U}\sum^{4}_{\ell}\Gamma^{2}_{\ell,i}\nonumber\\
&-&\Gamma_{1,i}\biggr(\sum^{N}_{\alpha=1}\Psi^{*}_{\alpha,i}\hat{\sigma}_{x}\Psi_{\alpha,i}\biggl)
+\Gamma_{2,i}\biggr(\sum^{N}_{\alpha=1}\Psi^{*}_{\alpha,i}\hat{\sigma}_{z}\Psi_{\alpha,i}\biggl)\nonumber\\
&-&\Gamma_{3,i}\biggr(\sum^{N}_{\alpha=1}K^{*}_{\alpha,i}\hat{\sigma}_{x}K_{\alpha,i}\biggl) 
+\Gamma_{4,i}\biggr(\sum^{N}_{\alpha=1}K^{*}_{\alpha,i}\hat{\sigma}_{z}K_{\alpha,i}\biggl)\biggl], \nonumber\\
\end{eqnarray}
where $\Gamma_{\ell,i}$ ($\ell =1,2,3,4,$) are the four scalar auxiliary fields. The scalar fields $\Gamma_{\ell,i}$ relate to the original fermion operator with the following relation:
\begin{eqnarray}
&&\frac{2}{U}\Gamma_{1,i}\longleftrightarrow  \Psi^{*}_{\alpha,i}\hat{\sigma}_{x}\Psi_{\alpha,i},\nonumber\\
&&\frac{2}{U}\Gamma_{2,i}\longleftrightarrow -\Psi^{*}_{\alpha,i}\hat{\sigma}_{z}\Psi_{\alpha,i},\nonumber\\
&&\frac{2}{U}\Gamma_{3,i}\longleftrightarrow K^{*}_{\alpha,i}\hat{\sigma}_{x}K_{\alpha,i},\nonumber\\
&&\frac{2}{U}\Gamma_{4,i}\longleftrightarrow -K^{*}_{\alpha,i}\hat{\sigma}_{z}K_{\alpha,i}.\nonumber
\end{eqnarray}
The above relations mean that $\Gamma_{1,i}$, $\Gamma_{2,i}$, $\Gamma_{3,i}$, and $\Gamma_{4,i}$ 
are related to the inner-bond order (the BI phase), the inner site density-wave order (the Aoki phase), the inter-bond order (the BDI-SPT phase), 
and the inter-site density-wave order (the Aoki phase), respectively. 
Actually, on the basis of the translational invariance in the i-SSH model, if we assume the semi-classical approximation and/or large-$N$ limit $N\to \infty$, 
the above relation can be exact, i.e., the arrow label in the above relations is replaced by equal sign. 

In addition, we comment that if one assumes the Schwinger fermion representation picture, 
the four scalar auxiliary fields $\Gamma_{\ell,i}$ can be mapped into the spin-$1/2$ variable defined on the {\it link} in the SSH model: 
$\frac{2}{U}\Gamma_{1,i}=S_{(i,i),x}$, $\frac{2}{U}\Gamma_{2,i}=S_{(i,i),z}$, $\frac{2}{U}\Gamma_{3,i}=S_{(i,i+1),x}$, and $\frac{2}{U}\Gamma_{4,i}=S_{(i,i+1),z}$, 
where $(i,i)$ is a link in unit cell $i$ and $(i+1,i)$ is a link between the $i$ and $i+1$ unit cells. 

In this work, our goal is to detect the global phase diagram. 
To this end, we apply mean-field treatment to the four scalar fields $\Gamma_{\ell,i}$. 
Because the system has discrete translational invariance for space, 
we drop the space and imaginary time dependence of $\Gamma_{\ell,i}$: $\Gamma_{\ell,i}(\tau)\rightarrow \Gamma_{\ell}$.
The partition function $Z$ including the mean-field $\Gamma_{\ell}$ can be written as
\begin{eqnarray}
&&Z=\int[dA^{*}_{\alpha,i}][dA_{\alpha,i} ][dB^{*}_{\alpha,i}][dB_{\alpha,i} ]\prod^{4}_{\ell=1}d\Gamma_{\ell}\nonumber\\
&&\times\exp\biggl[-S_{0}-\biggr(\frac{\beta N_{uc}N}{U}\sum^{4}_{\ell}\Gamma^{2}_{\ell}\biggl)\biggl],\nonumber\\
\end{eqnarray}
where, $N_{uc}$ is the number of unit cells in the periodic system, and the action $S_{0}$ is given by
\begin{eqnarray}
S_{0}&=&\int^{\beta}_{0}d\tau \biggl[ \sum_{i}\sum^{N}_{\alpha=1}\Psi^{*}_{\alpha,i}\partial_{\tau}\Psi_{i}+ \sum^{N}_{\alpha=1}h^{S}_{{\rm eff},\alpha}\biggr],\nonumber\\
h^{S}_{{\rm eff},\alpha}&=&\sum_{i}\biggr[(-J_{1}A^{*}_{\alpha,i}B_{\alpha,i}-J_{2}A^{*}_{\alpha, i+1}B_{\alpha,i}+\mbox{c.c.})\nonumber\\
&&-\Gamma_{1}(\Psi^{*}_{\alpha,i}\hat{\sigma}_{x}\Psi_{\alpha,i})+\Gamma_{2}(\Psi^{*}_{\alpha,i}\hat{\sigma}_{z}\Psi_{\alpha,i})\nonumber\\
&&-\Gamma_{3}(K^{*}_{\alpha,i}\hat{\sigma}_{x}K_{\alpha,i})
+\Gamma_{4}(K^{*}_{\alpha,i}\hat{\sigma}_{x}K_{\alpha,i})\biggl].
\end{eqnarray}
As seen from the action $S_{0}$, the interaction term has been modified into a bilinear form of the fermion field. 
Furthermore, since $\sum_{\alpha} h^{S}_{{\rm eff},\alpha}$ can be regarded as $N$-copies of the non-interacting Hamiltonian, 
we can drop the flavor index $\alpha$ and write the Hamiltonian as $\sum_{\alpha}h^{S}_{{\rm eff},\alpha}=Nh^{S}_{\rm eff}$.
Then, the Hamiltonian $h^{S}_{\rm eff}$ can be written as the following bulk-momentum representation
\begin{eqnarray}
h^{S}_{\rm eff}(\tau)&=&\sum_{k\in B.Z.}\Psi^{*}(k,\tau) h^{S}_{{\rm bulk}}(k)\Psi(k,\tau),\\
h^{S}_{\rm bulk}(k)&=&\biggr[-(J_{1}+\Gamma_{1})-(J_{2}+\Gamma_{3})\cos k\biggl]\hat{\sigma}_{x}\nonumber\\
&&+\biggr[-(J_{2}+\Gamma_{3})\sin k\biggl]\hat{\sigma}_{y}+(\Gamma_{2}+\Gamma_{4})\hat{\sigma}_{z},\nonumber\\
\label{BkSSH}
\end{eqnarray}
where B.Z. means the first Brillouin Zone. Here, please note that, if for the Hamiltonian $h^{S}_{\rm bulk}$ 
we take $\Gamma_{3}=0$ and $\Gamma_{4}=0$, it ends up being dealing with the type-I i-SSH model. 
Therefore, the effective bulk momentum Hamiltonian $h^{\rm I}_{\alpha}(k)$ in the type-I i-SSH model in the main text can be obtained.
The bulk momentum spectrum $E_{\pm}(k)$ of the Hamiltonian $h^{S}_{\rm bulk}$ is given by
\begin{eqnarray}
&&E_{\pm}(k,\Gamma_{\ell})=\pm\biggr[\biggr( -(J_{1}+\Gamma_{1})-(J_{2}+\Gamma_{3})\cos k\biggl)^{2} \nonumber\\
&&+\biggr( -(J_{2}+\Gamma_{3})\sin k \biggl)^{2} + (\Gamma_{2}+\Gamma_{4})^{2}\biggl]^{1/2}.
\end{eqnarray}
This spectrum will be used later. 

Here, we should comment that the bulk-momentum Hamiltonian $h^{S}_{\rm bulk}(k)$ of Eq.~(\ref{BkSSH}) 
belongs to the BDI class in the classification theory of the non-interacting topological Hamiltonian \cite{Schnyder,Ryu,Kitaev} 
if $\Gamma_{2}, \:\Gamma_{4}=0$. Conversely, for a finite case, $\Gamma_{2}, \:\Gamma_{4}\neq 0$, 
the Hamiltonian is no longer BDI class because the $\hat{\sigma}_{z}$ term in $h^{S}_{\rm bulk}(k)$ breaks the chiral symmetry: 
$\hat{\sigma}_{z} h^{S}_{\rm bulk}(k)\hat{\sigma}_{z}\neq -h^{S}_{\rm bulk}(k)$.  
Therefore, if there is a mean-field solution with $\Gamma_{2},\:\Gamma_{4}\neq 0$, the system is not BDI class. 
This fact means that the system does not possess a topological non-trivial phase.

Also, if we define the effective hopping parameter as $\tilde{J}_{1}\equiv J_{1}+\Gamma_{1}$, and $\tilde{J}_{2}\equiv J_{2}+\Gamma_{3}$, 
then intuitively, $\Gamma_{1}$ and $\Gamma_{3}$ can be regarded as a correction effect for the bear hopping parameter. 
Since the values of $\Gamma_{1}$ and $\Gamma_{3}$ depend on the value of $U$, 
in the large-$N$ formalism, the interaction $U$ changes the hopping strength effectively. 

The present action $S_{0}$ is a quadratic form of the fermion fields. Thus, we can integrate out the fermion fields. 
We can obtain $N$-copies of the effective action represented by only the mean-fields $\Gamma_{\ell}$. 
The effective action $S_{\rm eff}$ is given as 
\begin{eqnarray}
Z&=&\int\prod^{4}_{\ell=1}d\Gamma_{\ell}e^{-N S_{\rm eff}},\nonumber\\
S_{\rm eff}&=&\beta N_{uc}\int \int \frac{d\omega d k}{4\pi^{2}}\nonumber\\
&&\times\biggr[\frac{1}{U}\sum^{4}_{\ell}\Gamma^{2}_{\ell} -\log[\omega^{2}+E^{2}_{+}(k,\Gamma_{\ell})]\biggl],
 \label{GL}
\end{eqnarray}
where the imaginary time $\tau$ has been replaced by the Matsubara frequency $\omega$. 
Because we consider the zero-temperature limit here, $\omega$ can be treated as a continuous variable. 
Thus, the integral of $\omega$ appears in the effective action $S_{\rm eff}$. 

Let us assume the large-$N$ limit, $N\to \infty$. The groundstate phase diagram at zero temperature 
can be calculated by solving the saddle point equations of the action $S_{\rm eff}$.
In the large-$N$ limit, the mean-field solutions of $\Gamma_{\ell}$ are believed to be exact \cite{Gross-Neveu,Auerbach}. 
We denote the solutions of $\Gamma_{\ell}$ by $\Gamma_{\ell,0}$. The saddle point equations are given by 
\begin{eqnarray}
&&\left.\frac{\partial S_{\rm eff}}{\partial \Gamma_{\ell}} \right|_{\Gamma_{\ell}=\Gamma_{\ell,0}}=0\nonumber\\
\Rightarrow && \frac{\Gamma_{\ell,0}}{U}
=\int^{\pi}_{-\pi}\frac{dk}{4\pi} \left.\biggr( \frac{\partial E_{+}(k,\Gamma_{\ell}) }{\partial\Gamma_{\ell}} \biggl)\right|_{\Gamma_{\ell}=\Gamma_{\ell,0}},
\label{gap}
\end{eqnarray}
where we have integrated out the variable $\omega$ in the saddle point equation. The obtained equation of Eq.~(\ref{gap}) is just the gap equation. 
Thus, the solutions of $\Gamma_{\ell,0}$ and the global phase diagram can be obtained in a numerical self-consistent way. 
Also, this gap equation covers the type-I and type-II i-SSH model. Therefore we can obtain the groundstate phase diagram for both the type-I and type-II i-SSH model.
The phase diagram in $J_{1}$-$U$ parameter space is shown in Fig.~1 in the main text. 
 
\begin{figure}[t]
\begin{center} 
\includegraphics[width=5cm]{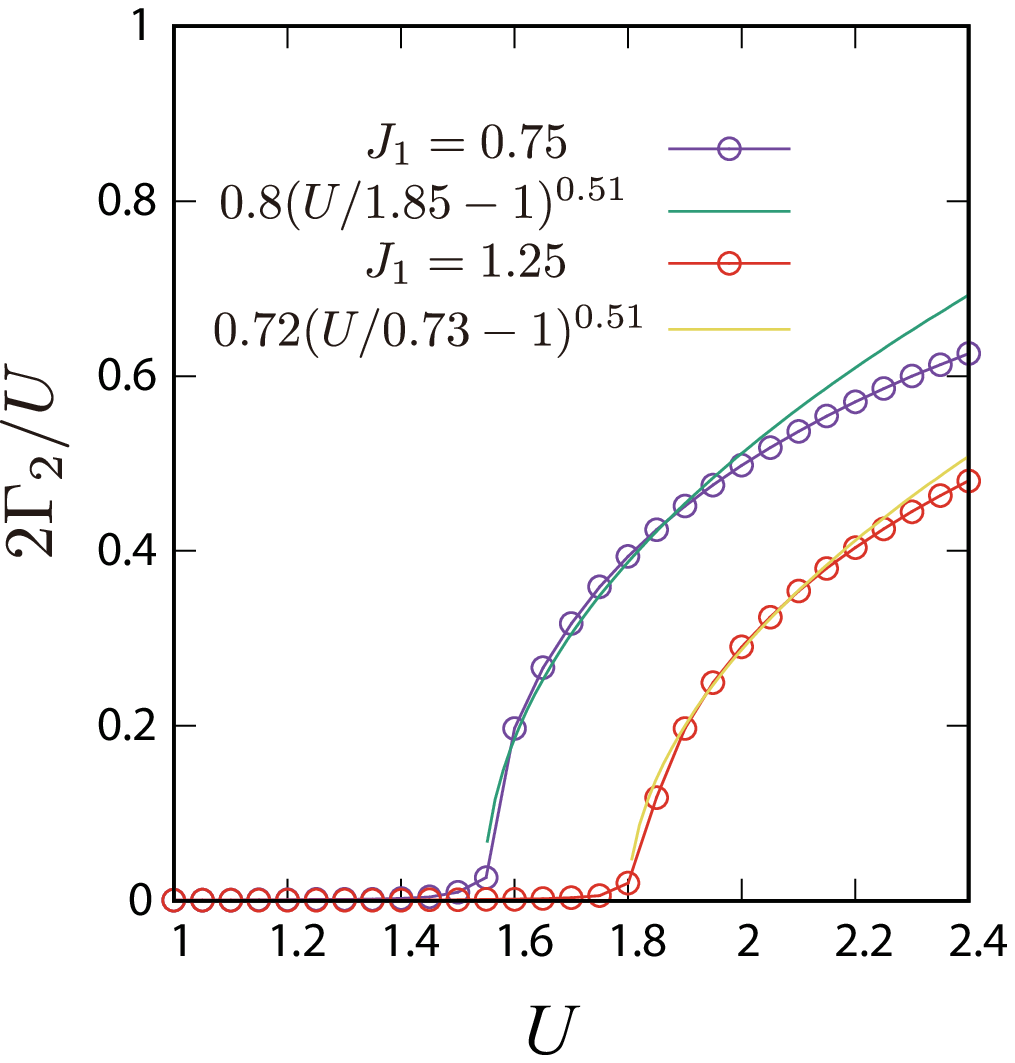}
\end{center} 
\caption{The typical behavior of $2\Gamma_{2}/U$ with increasing $U$ for $J_{1}=0.75$ and $J_{1}=1.25$ with $J_{2}=1$. 
The calculated data lines lie on a scaling function around the phase transition point.
In the phase transition to the Aoki phase, the value of the critical exponent is ${\bar \beta}=0.51$.}
\label{Critical}
\end{figure}
\subsection*{Critical behavior of $\Gamma_{2}$ }
We investigate the critical behavior of the solution $\Gamma_{2}$. 
We plot $2\Gamma_{2}/U$ with increasing $U$ at $J_{1}/J_{2}=0.75$ and $J_{1}/J_{2}=1.25$. 
The result is shown in Fig.~\ref{Critical}. 
As seen from the result, the phase transition to the Aoki phase is a continuous second-order type. 
In addition, the critical exponent ${\bar \beta}$, defined as $2\Gamma_{2(4)}/U \propto |U/U_{c}-1|^{{\bar \beta}} $ ($U_{c}$ is a transition point), is extracted from a data fitting. 
For both $J_{1}/J_{2}$ cases in Fig.~\ref{Critical}, we obtain the same value of the critical exponent ${\bar \beta}\sim 0.51$. 
The obtained value of ${\bar \beta}$ is much close to the pure mean-field value ${\bar \beta}=0.5$, expected in the CGNW model \cite{Aoki}. 
Both from the BI phase to the Aoki phase and the BDI-SPT phase to the Aoki phase, the same critical phenomena are expected in the large-$N$ case. 
Also, we confirmed that $\Gamma_{2}$ and $\Gamma_{4}$ have the same behavior in our numerical calculation.   
\begin{figure}[t]
\begin{center} 
\includegraphics[width=8cm]{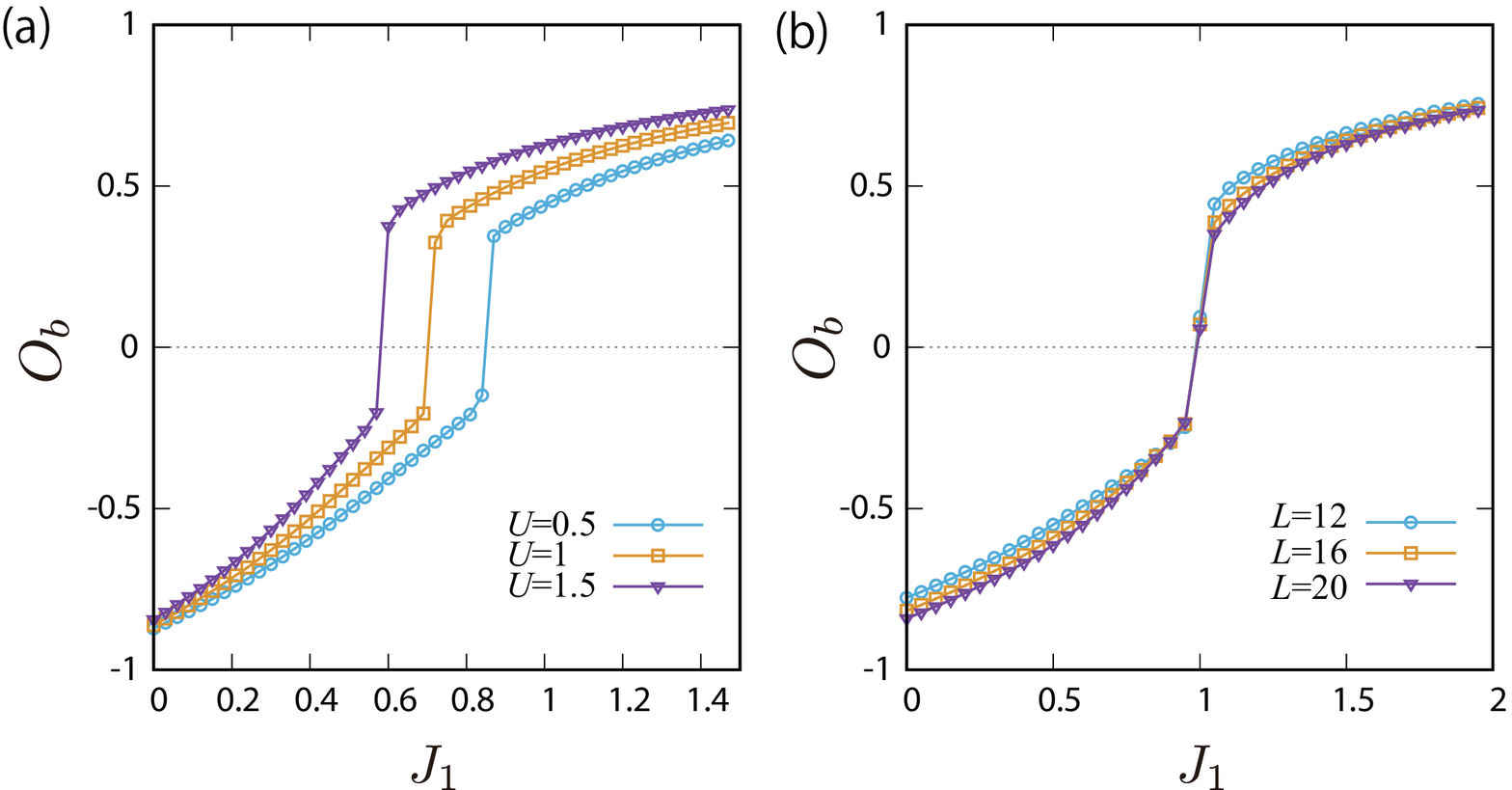}
\end{center} 
\caption{The behavior of $O_{b}$: 
(a) $U$ and $J_{1}$ dependence in the type-I i-SSH model with $L=16$. 
(b) System size and $J_{1}$ dependence in the type-II i-SSH model. 
For both cases, $J_{2}=1$.}
\label{bond}
\end{figure}

\section*{B. Supplemental result in exact diagonalization}

\begin{figure}[t]
\begin{center} 
\includegraphics[width=8cm]{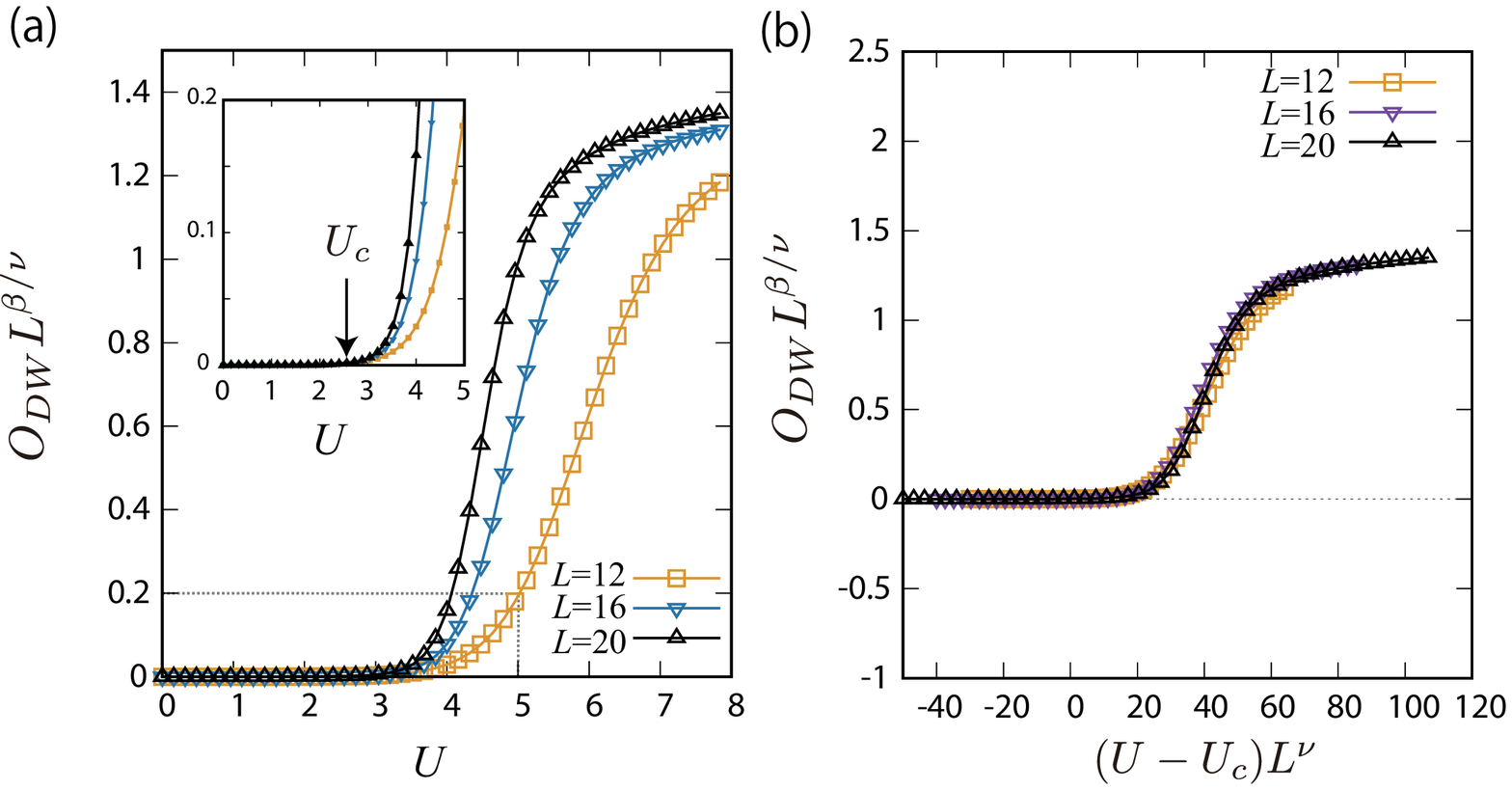}
\end{center} 
\caption{Critical behavior to the Aoki phase in the type-I i-SSH model. 
The typical behavior of $O_{DW}$ with increasing $U$ for $J_{1}/J_{2}=0$.
The data are fitted by $d=2$ Ising-type critical exponent, $\beta=0.125$, $\nu=1$. We obtain $U_{c}= 2.55$}
\label{Critical2}
\end{figure}

\begin{figure}[t]
\begin{center} 
\includegraphics[width=8cm]{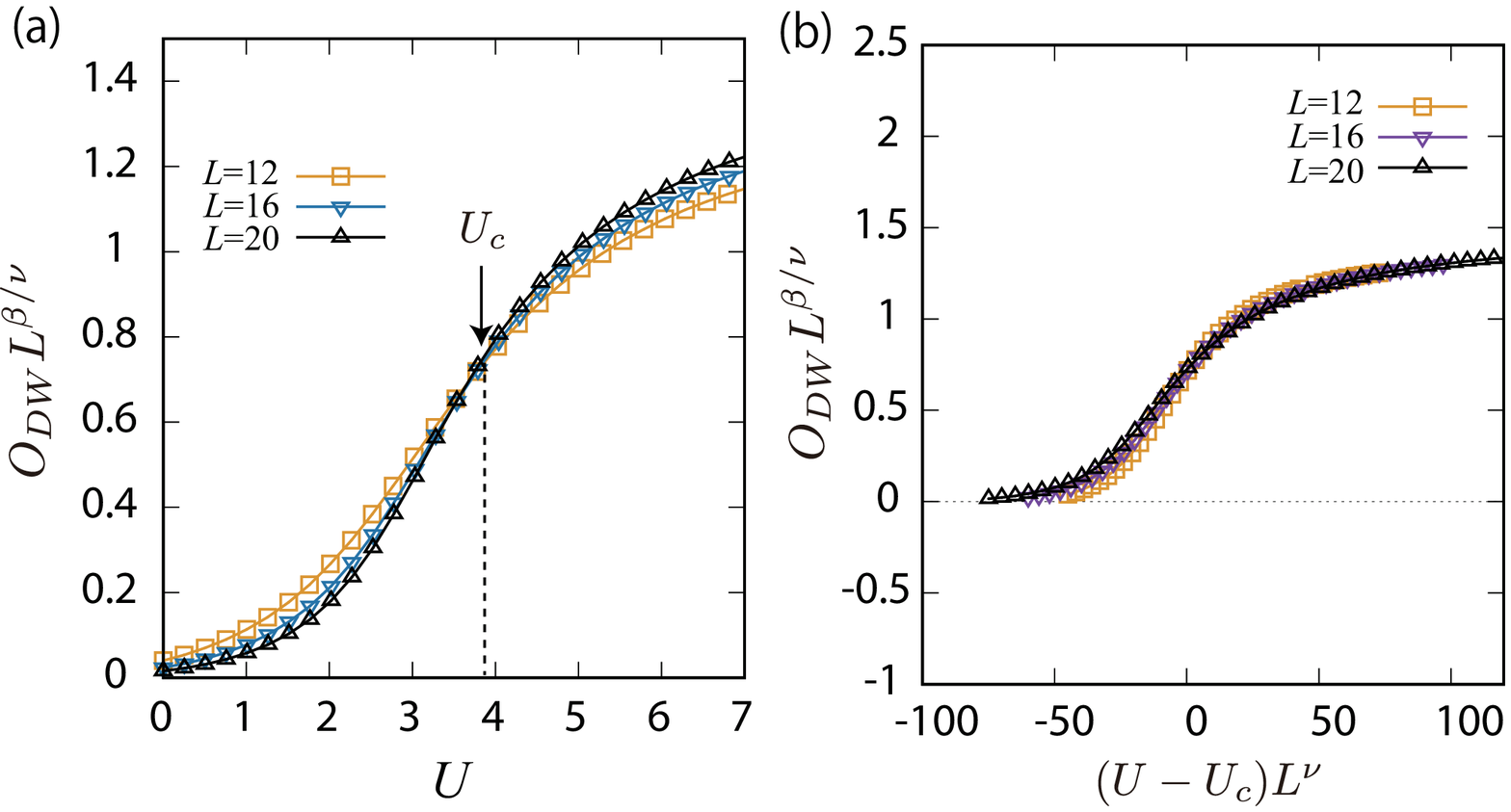}
\end{center} 
\caption{Critical behavior to the Aoki phase. 
The typical behavior of $O_{DW}$ with increasing $U$ for $J_{1}/J_{2}=1$.
The data are fitted by $d=2$ Ising-type critical exponent, $\beta=0.125$, $\nu=1$. We obtain $U_{c}= 3.85$}
\label{Critical1}
\end{figure}
In the exact diagonalization, the system with $L=12, 16$, and $20$ at half-filling is employed. 
Practically, we have used the Lanczos algorithm \cite{EDtext1,EDtext2} to obtain groundstate wave functions. 
Using finite-size scaling, we can determine the phase boundary and the critical phenomena in the thermodynamic limit. 
To calculate the groundstate phase diagram, we introduce the difference of the inner- and inter-bond order parameter $O_{b}$, and the density-wave order parameter $O_{DW}$ as 
\begin{eqnarray}
&&O_{b}=\frac{1}{L}\sum_{i}\biggl[\langle a^{\dagger}_{i}b_{i}+\mbox{h.c.} \rangle -\langle a^{\dagger}_{i+1}b_{i}+\mbox{h.c.}\rangle\biggr],\\
&&O_{DW}=\frac{1}{L}\sum_{i}\biggl[\langle |a^{\dagger}_{i}a_{i}-b^{\dagger}_{i}b_{i}|^{2}\rangle+\langle |a^{\dagger}_{i+1}a_{i+1}-b^{\dagger}_{i}b_{i}|^{2}\rangle \biggr],\nonumber\\
\label{OP}
\end{eqnarray}
where $\langle \cdot \rangle$ means the groundstate expectation value, the flavor index of the fermion operator $\alpha$ has been omitted because the $N=1$ case is considered, 
and $L$ is the number of lattice sites in the periodic system satisfying $L=2N_{us}$.
If $O_{b}>0$ ($<0$) and $O_{DW}=0$, the system is in the BI phase (the BDI-SPT phase). If $O_{DW}>0$, the Aoki phase appears.
Using the order parameters, we obtain the groundstate phase diagrams in Fig.~2 in the main text. 
Here, we show the detailed behavior of $O_{b}$ and $O_{DW}$, their system size dependence, and the scaling behavior.
Figure~\ref{bond} (a) and (b) display the $U$ dependence of $O_{b}$ in the type-I i-SSH model 
and the system size dependence of $O_{b}$ in the type-II i-SSH model with varying $J_{1}$ and $J_{2}=1$. 
For both models, the clear phase transition between the BI and the BDI-SPT phase is captured.
In Fig.~\ref{bond} (a), the transition point between the BI and the BDI-SPT phase is shifted with increasing $U$; 
this behavior is the same as that in the large-$N$ expansion. 
Also in Fig.~\ref{bond} (b), the result indicates no system-size dependence. 
Therefore, we can easily determine the phase boundary between the BI and the BDI-SPT phase. 
 
Next, we focus on the phase transition toward the Aoki phase. 
In our exact diagonalization, the behavior of $O_{DW}$ exhibits continuous behavior and explicit system-size dependence. 
Therefore, to determine the phase boundary of the Aoki phase in the type-I and type-II i-SSH model, finite-size scaling is used. 
In a previous study \cite{Bermudez}, the critical behavior toward the Aoki phase in the CGNW model was investigated using a matrix-product state numerical simulation. 
The numerical study \cite{Bermudez} indicates that the phase transition toward the Aoki phase belongs to the $d=2$ Ising-type universality class. 
With respect to the result in \cite{Bermudez}, we also carry out finite-size scaling for the type-I and type-II i-SSH model 
and investigate whether the critical behavior in the type-I and type-II i-SSH model belongs to the $d=2$ Ising-type universality class. 
Figure~\ref{Critical2} is the typical result of the finite-size scaling of $O_{DW}$ in the type-I SSH model with $J_{1}=0$ and $J_{2}=1$, 
where the phase transition from the BDI-SPT to the Aoki phase occurs. 
Figure~\ref{Critical2} (a) shows the behavior of $O_{DW}L^{\beta/\nu}$ along the $U$-axis, where $\beta$ and $\nu$ are critical exponents.
Here, we find that suitable choice are $\beta=0.125$ and $\nu=1$; 
then, three pieces of data with different system sizes start to separate at one point, which can be regarded as transition point $U_{c}$. 
Therefore, we obtain the transition point $U_{c}=2.55$. Actually, using the obtained value of $U_{c}$, the three pieces of data are also plotted along the axis $(U-U_{c})L^{\nu}$. Here, we consider the scaling ansatz defined by $\Psi[(U-U_{c})L^{1/\nu}]=L^{\beta/\nu}O_{DW}$, where $\Psi$ is a scaling function. 
If we choose the correct values of $\beta$ and $\nu$, the three pieces of data must overlap. As shown in Fig.~\ref{Critical2} (b), the data do almost overlap. 
This result indicates that the universality class of the phase transition is the same as that of the CGNW model, which is the $d=2$ Ising-type $\beta=0.125$ and $\nu=1$. 
Furthermore, we carry out the same scaling procedure for the phase transition to the Aoki phase in the type-II i-SSH model. 
The results for $J_{1}=1$ and $J_{2}=1$ are displayed in Fig~\ref{Critical1} (a) and (b). 
In Fig.~\ref{Critical1} (a), we find that for $\beta = 0.125$ and $\nu = 1$, three pieces of data with different system sizes intersect at one point. 
That is, the transition point $U_{c}=3.85$ is obtained. This value is fairly close to that obtained in a previous study \cite{Sirker}. 
Also, we obtain a clear overlap of the three data points, as shown in Fig.~\ref{Critical1} (b).
Therefore, the phase transition also belongs to the $d=2$ Ising-type universality class. 

\section*{C. Concrete setup in an optical lattice}

\subsection*{Type-I i-SSH model}
In the main text, we proposed an implementation scheme of the type-I i-SSH model with $N=1$.
For the implementation scheme, we show a concrete setup by employing a ${}^{173}{\rm Yb}$ cold-atom gas in an optical lattice \cite{Taie}. 
$^{173}$Yb atom has six different internal states (different nuclear spin: $I=\pm1/2$, $\pm 3/2$, $\pm 5/2$). In this proposal, two of their internal states are selected.
As a typical feature of ${}^{173}{\rm Yb}$, the s-wave scattering lengths between each internal states are finite and equivalent \cite{Taie}. 
Therefore, on-site interactions between the two different internal states exist.
As shown in Fig.~3 (a) in the main text, two different kinds of one-dimensional double-well optical lattice are prepared. 
Each internal state of atoms is trapped in each double-well optical lattice.
These two different double-well optical lattice may be created using a spin-dependent optical lattice technique that originates from a vector-light shift \cite{SPL1,SPL2}.
Then, the two different double-well potentials are given by $V_{1}(x)=-V_{s}\sin^{2}(2kx)-V_{l}\cos^{2}(kx)$ and $V_{2}(x)=-V_{s}\sin^{2}(2kx+\phi_{s})-V_{l}\cos^{2}(kx+\phi_{l})$, where $V_{s(l)}$ is the lattice depth for short (long) lattice potential with $k=2\pi/\lambda$.
$\lambda$ is the wave-length of the long lattice potential. 
The lattice potential creates $\lambda/4$ lattice spacing. 
Let $\lambda/2$ be the unit length. If we set $\phi_{s}=\pi$ and $\phi_{l}=\pi/2$, 
the desired one-dimensional double-well optical lattices shown in Fig.~3 (a) in the main text can be created. 

When we set the lattice potential $V_{1}(x)$ and $V_{2}(x)$ deep, the Wannier function can be introduced around each potential minimum. 
Then, an on-site interaction between the two different internal states exists on same site. By using the Wannier function, this interaction can be written as 
\begin{eqnarray}
U=\frac{4\pi\hbar^{2}a_{s}}{m}\int d{\bf r}|w_{i,L(R)}({\bf r})|^{4}, \label{Urep}
\end{eqnarray}
where $a_{s}$ and $m$ are the s-wave scattering length and atom mass, respectively, 
$w_{i,L(R)}({\bf r})=w^{x}_{i,L(R)}(x)w_{i,y}(y)w_{i,z}(z)$, $w_{i,L(R)}({\bf r})$ is a left (right) lowest-band Wannier function spanned on the unit-cell site $i$ in lowest band, and $w^{x}_{i,L(R)}(x)$ is a left (right) site Wannier function in the $x-$ direction double-well optical lattice.
$w_{y,i}(y)$ and $w_{z,i}(z)$ are $y$- and $z$-direction Wannier functions to confine atoms on a one-dimensional line ($x$-direction).  
Here, we can estimate the parameters in the type-I i-SSH model in a concrete set of lattice parameters.
With respect to a real experimental system \cite{Taie}, we set $a_{s}=10.55$[nm], $\lambda=1064$[nm], and $V_{l}=10E_{R}$, and take energy unit $E_{R}=h^{2}/2m\lambda^{2}$.  
$y$- and $z$-direction confinement lattice potentials are set with $\lambda/4$ lattice spacing and the potential depth $V_{y}=20E_{R}$, $V_{z}=20E_{R}$. 

\begin{figure}[t]
\begin{center} 
\includegraphics[width=8cm]{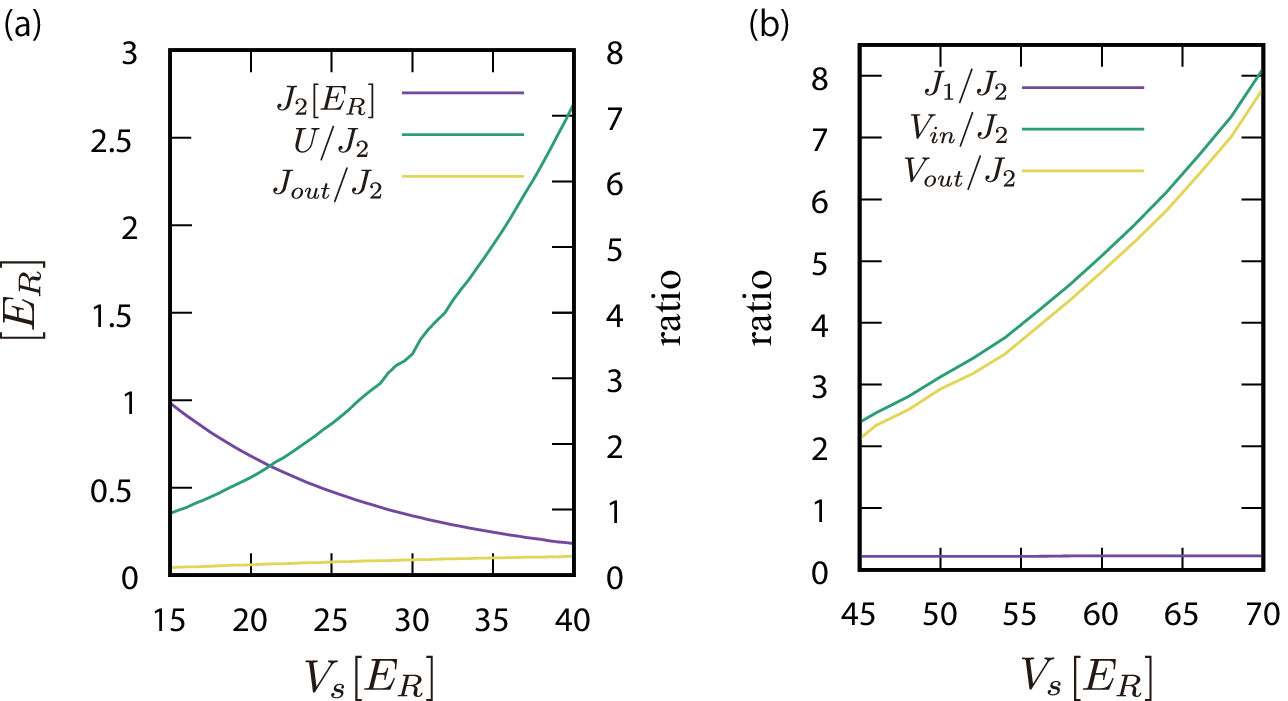}
\end{center} 
\caption{(a) The behavior of $J_2 [E_R]$, $U/J_{2}$, and $J_{out}/J_{2}$ in the type-I i-SSH model. 
(b) The behavior of $V_{in}/J_2$, $V_{out}/J_{2}$, and $J_{1}/J_{2}$ in the type-I i-SSH model.}
\label{Hpara}
\end{figure}

As shown in Fig.~\ref{Hpara} (a), we calculate the $V_{s}$ dependence of the SSH parameters $J_2 [E_R]$, $U/J_{2}$.
Also, we calculate $J_{out}/J_{2}$, where $J_{out}$ is the hopping amplitude between the NN unit-cells shown in Fig.~3 (a).
The result indicates that $J_{out}$ is adequately suppressed and the value of $U/J_{2}$ covers our target parameter regime in the main text. 
Thus, if $\Omega$ is widely controllable, we expect that the experimental system can cover our target parameter regime for the type-I i-SSH model in the main text.

\subsection*{Type-II i-SSH model}
For the type-II i-SSH model, assuming concrete experimental parameters, 
we also estimate the values of the parameters of the type-II i-SSH model.
As mentioned in the main text a dipolar fermionic atom is employed. 
As a concrete example, we consider $^{167}$Er. 
This atom has a large magnetic dipole moment $\mu=7\mu_{B}$  
($\mu_{B}$ is the Bohr magneton). 
The double-well optical lattice potential suggested in Fig.~3 (b) in the main text is given by $V_{1}(x)=-V_{s}\sin^{2}(2kx)-V_{l}\cos^{2}(kx)$ with $k=2\pi/\lambda$ ($\lambda=1063$ [nm]). 
The lattice spacing is $\lambda/4$. Then, using the Wannier functions as in Eq.~(\ref{Urep}), 
the dipole-dipole interaction (DDI) between nearest-neighbor sites \cite{Dutta} in the system can be written as follows:
\begin{eqnarray}
V_{in}&=&D\int d{\bf r}d{\bf r}'\nonumber\\
&&\times|w_{i,L}({\bf r})|^{2}\biggl[\frac{1-3\cos^{2}\theta_{{\bf r}-{\bf r}'}}{|{\bf r}-{\bf r}'|^{3}}\biggr]|w_{i,R}({\bf r}')|^{2},\nonumber\\
V_{out}&=&D\int d{\bf r}d{\bf r}'\nonumber\\
&&\times|w_{i,R}({\bf r})|^{2}\biggl[\frac{1-3\cos^{2}\theta_{{\bf r}-{\bf r}'}}{|{\bf r}-{\bf r}'|^{3}}\biggr]|w_{i+1,L}({\bf r}')|^{2},\nonumber
\end{eqnarray}
where $V_{in}$ is the DDI between the NN sites in same unit cell $i$ and $V_{out}$ is the DDI between the NN sites in the NN unit cells ($i$ and $i+1$). 
${\bf r}=(x,y,z)$ is scaled by $\lambda$ and $\cos\theta_{{\bf r}-{\bf r}'}=(z-z')/|{\bf r}-{\bf r}'|$. 
$D=\frac{\mu_{0}\mu^{2}m}{\pi^{3}\hbar^{2}\lambda }$[$E_{R}$], $\mu_{0}$ is the vacuum permeability.
We calculate the values of the parameters of the type-II i-SSH model by taking an energy unit $E_{R}=h^{2}/2m\lambda^{2}$. 
Here, $y$- and $z$-direction confinement lattice potentials are set with lattice spacing $\lambda/4$ and potential depth $V_{y}=20E_{R}$, $V_{z}=20E_{R}$, and the long lattice depth in the $x$-direction double-well lattice potential is set as $V_{l}=2E_{R}$. 
In Fig.~\ref{Hpara} (b), the $V_{s}$ dependence of the parameters $J_{1}/J_{2}$, $V_{in}/J_{2}$, and $V_{out}/J_{2}$ is plotted. 
The result indicates that for the small $J_{1}/J_{2}$ regime, $V_{in}/J_{2}$ and $V_{out}/J_{2}$ are adequately large for the phase transition to the Aoki phase to occur in the main text.
In our estimation, the values of the two interactions $V_{in}$ and $V_{out}$ are close to each other. Thus, $U$ in the type-II i-SSH model can be approximated as $U=(V_{in}+V_{out})/2$.


\begin{thebibliography}{99}
\bibitem{Shen}
S.-Q. Shen, \textit{Topological Insulators} (Springer-Verlag, Berlin, 2012).

\bibitem{Fradkin}
E. Fradkin, Field Theories of Condensed Matter Physics (Cambridge University Press 2013).

\bibitem{Wilson}
K. G. Wilson, New Phenomena in Subnuclear Physics (Erice, 1975), (Plenum, New York, 1977).

\bibitem{Rothe}
H. J. Rothe, World Sci. Lect. Notes Phys. {\bf 82},1 (2012).

\bibitem{Creutz2}
M. Creutz, Rev. Mod. Phys. {\bf 73}, 119 (2001).

\bibitem{Bermudez}
A. Bermudez, E. Tirrito, M. Rizzi, M. Lewenstein, and S. Hands, Annals of Physics {\bf 399}, 149 (2018).

\bibitem{Cirac}
J. I. Cirac, P. Maraner, and J. K. Pachos, Phys. Rev. Lett. {\bf 105}, 190403 (2010).

\bibitem{Zache}
T. V. Zache, F. Hebenstreit, F. Jendrzejewski, M. K. Oberthaler, J. Berges, and P. Hauke, 
Quantum Sci. Technol. {\bf 3}, 034010 (2018).

\bibitem{Kuno}
Y. Kuno, I. Ichinose, and Y. Takahashi, Sci. Rep. {\bf 8}, 10699 (2018).

\bibitem{Aoki}
S. Aoki, Phys. Rev. D {\bf 30}, 2653 (1984). 

\bibitem{Araki}
Y. Araki and T. Kimura, Phys. Rev. B {\bf 87}, 205440 (2013); 
Y. Araki, T. Kimura, A. Sekine, K. Nomura, and T. Z. Nakano. arXiv:1311.3973.

\bibitem{SSH}
A. J. Heeger, S. Kivelson, J. R. Schrieffer, and W. P. Su, Rev. Mod. Phys. {\bf 60}, 781 (1988).

\bibitem{Asboth}
J. K. Asboth, L. Oroszlany, and A. Palyi, A Short Course on Topological Insulators (Springer International Publishing, New York, 2016), Vol. 919.

\bibitem{Creutz}
M. Creutz, T. Kimura, and T. Misumi, Phys. Rev. D {\bf 83}, 094506 (2011).

\bibitem{Gross-Neveu}
D. J. Gross and A. Neveu, Phys. Rev. D {\bf 10}, 3235 (1974). 

\bibitem{Cazalilla2} 
M. A. Cazalilla and A. M. Rey, Rep. Prog. Phys. {\bf 77}, 124401 (2014).

\bibitem{Taie}
S. Taie, R. Yamazaki, S. Sugawa, and Y. Takahashi, Nat. Phys. {\bf 8}, 825 (2012).

\bibitem{Gorshkov}
A. V. Gorshkov, M. Hermele, V. Gurarie, C. Xu, P. S. Julienne, J. Ye, P. Zoller, E. Demler, M. D. Lukin, and A. M. Rey, Nat. Phys. {\bf 6}, 289 (2010).

\bibitem{Inouye}
S. Inouye, M. R. Andrews, J. Stenger, H.-J. Miesner, D. M. Stamper-Kurn, and W. Ketterle, Nature {\bf 392}, 151 (1998).

\bibitem{Hofer}
M. Hofer, L. Riegger, F. Scazza, C. Hofrichter, D. R. Fernandes, M. M. Parish, J. Levinsen, I. Bloch, and S. Folling, Phys.Rev.Lett. {\bf 115} 265302 (2015).

\bibitem{Ferlaino}
S. Baier, M. J. Mark, D. Petter, K. Aikawa, L. Chomaz, Z. Cai, M. Baranov, P. Zoller, and F. Ferlaino, Science {\bf 352}, 201 (2016).

\bibitem{interaction}
In experiments, the Feshbach resonance technique can make on-site interactions attractive. If one considers a dipolar atom that generates a repulsive NN DDI, the amplitude of the attractive interaction can be tuned to be same order as that of the DDI.

\bibitem{Affleck}
I. Affleck and F. D. M. Haldane, Phys. Rev. B  {\bf 36}, 5291 (1987);
J. B. Marston and I. Affleck, Phys. Rev. B {\bf 39}, 11538 (1989).

\bibitem{CChiral}
The continuous chiral symmetry operation is defined by 
$\psi_{\alpha,i}\to e^{i\theta \gamma^{5}}\psi_{\alpha,i}$ and $\bar{\psi}_{\alpha,i}\to \bar{\psi}_{\alpha,i}e^{i\theta \gamma^{5}}$, where $\theta$ is an arbitrary phase parameter.

\bibitem{Fuertes}
W. G. Fuertes and J. M. Guilarte, J. Math. Phys. {\bf 38} 6214 (1997).

\bibitem{Reisz}
T. Reisz, Lect. Notes Phys. {\bf 508}, 192 (1998).

\bibitem{Sup}
See Supplemental Material for detailed technical aspects, which includes 
Refs. \cite{Auerbach,SPL2,Dutta}.
\bibitem{Auerbach}
A. Auerbach, Interacting Electrons and Quantum Magnetism, Graduate Texts in Contemporary Physics (Springer New York, 2012).

\bibitem{SPL2}
B. Yang, H. N. Dai, H. Sun, A. Reingruber, Z. S. Yuan, and J. W. Pan, Phys. Rev. A {\bf 96}, 011602(R) (2017).

\bibitem{Dutta}
O. Dutta, M. Gajda, P. Hauke, M. Lewenstein, D.-S. Luhmann, B. A. Malomed, T. Sowinski, and J. Zakrzewski, Rep. Prog. Phys. {\bf 78}, 066001 (2015).

\bibitem{Schnyder}
A. P. Schnyder, S. Ryu, A. Furusaki, and A.W.W. Ludwig, Phys. Rev. B {\bf 78}, 195125 (2008).

\bibitem{Ryu}
S. Ryu, A. P. Schnyder, A. Furusaki, and A. W. W. Ludwig, New J. Phys. {\bf 12}, 065010 (2010).

\bibitem{Kitaev}
A. Kitaev, AIP Conf. Proc. {\bf 1134}, 22 (2009)

\bibitem{Chiral}
The definition of these symmetries is that a Hamiltonian $H(\bf{k})$ possesses the following conditions:
for $S$ symmetry $\Gamma H(\bf{k})\Gamma^{-1}\neq -H(\bf{k})$, 
for $T$ symmetry $U_{T}^{-1}H(\bf{k})U_{T}\neq H(\bf{-k})$, and 
for $C$ symmetry $U_{C}^{-1}H(\bf{k})U_{C}\neq -H(\bf{-k})$, 
where $\Gamma$, $U_{T}$, and $U_{C}$ are some symmetry operators and, $U_{T}^{2} = +1$ and $U_{C}^{2} = +1$ are satisfied \cite{Schnyder}.
For $h^{S}_{\alpha}(k)$, the operators are given by $\Gamma=\hat{\sigma}_{z}$, $U_{T}=K$, and $U_{C}=\hat{\sigma}_{z}K$, where $K$ is an imaginary conjugation operator. It is noted that the definition of the $S$ symmetry used here is different from that used in high-energy physics \cite{CChiral}.

\bibitem{Senthil}
T. Senthil, Annual Review of Condensed Matter Physics {\bf 6}, 299 (2015).

\bibitem{Sirker}
J. Sirker, M. Maiti, N. P. Konstantinidis, and N. Sedlmayr, J. Stat. Mech. P10032 (2014).

\bibitem{EDtext1}
P. Prelovsek, J. Bonca, Strongly Correlated Systems: Numerical Methods, vol. {\bf 176}, Springer, 2013.

\bibitem{EDtext2}
M. Noack, S. R. Manmana, AIP Conf. Proc. {\bf 789}, 93-163 (2005).

\bibitem{scaling}
In our numerics, the scaling ansatz is $\Psi[(U-U_{c})L^{1/\nu}]=L^{\beta/\nu}O_{DW}$, where $\Psi$ is a scaling function and $U_{c}$ is transition point, and $\beta$ and $\nu$ are critical exponents. We used the three data sets from different system sizes $L=12$, $16$, and $20$.


\bibitem{Atala}
M. Atala, M. Aidelsburger, J.T. Barreiro, D. Abanin, T. Kitagawa, E. Demler, and I. Bloch, Nat. Phys. {\bf 9}, 795 (2013).

\bibitem{Gadway}
E. J. Meier, F. A. An, and B. Gadway, Nat. Commun. {\bf 7}, 13986 (2016).

\bibitem{Song}
B. Song, L. Zhang, C. He, T.F.J. Poon, E. Hajiyev, S. Zhang, X.-J. Liu, and G.-B. Jo, Sci. Adv. {\bf 4}, eaao4748 (2018).

\bibitem{SPL1}
P. Soltan-Panahi, J. Struck, P. Hauke, A. Bick, W. Plenkers, G. Meineke, C. Becker, P. Windpassinger, M. Lewenstein, and K. Sengstock, Nat. Phys. {\bf 7}, 434 (2011).

\bibitem{synthetic}
This situation corresponds to a synthetic dimensional technique. 
For its experimental realization, please refer to 
M. Mancini, et.al., Science {\bf 349}, 1510 (2015) and a. Celi, et. al. , Phys. Rev. Lett. {\bf 112}, 043001 (2014).

\bibitem{Lev}
M. Lu, N. Q. Burdick, and B. L. Lev, Phys. Rev. Lett. {\bf 108}, 215301 (2012).

\end{thebibliography}

\begin{thebibliography}{99}
\bibitem{Schnyder}
A. P. Schnyder, S. Ryu, A. Furusaki, and A.W.W. Ludwig, Phys. Rev. B {\bf 78}, 195125 (2008).

\bibitem{Ryu}
S. Ryu, A. P. Schnyder, A. Furusaki, and A. W. W. Ludwig, New J. Phys. {\bf 12}, 065010 (2010).

\bibitem{Kitaev}
A. Kitaev, Periodic table for topological insulators and superconductors, AIP Conf. Proc. {\bf 1134}, 22 (2009).

\bibitem{Gross-Neveu}
D. J. Gross and A. Neveu, Phys. Rev. D {\bf 10}, 3235 (1974). 

\bibitem{Auerbach}
A. Auerbach, Interacting Electrons and Quantum Magnetism, Graduate Texts in Contemporary Physics (Springer New York, 2012).

\bibitem{Aoki}
S. Aoki, Phys. Rev. D {\bf 30}, 2653 (1984). 

\bibitem{EDtext1}
P. Prelovsek, J. Bonca, Strongly Correlated Systems: Numerical Methods, vol. {\bf 176}, Springer, 2013.

\bibitem{EDtext2}
M. Noack, S. R. Manmana, AIP Conf. Proc. {\bf 789}, 93-163 (2005).

\bibitem{Bermudez}
A. Bermudez, E. Tirrito, M. Rizzi, M. Lewenstein, and S. Hands, Annals of Physics {\bf 399}, 149 (2018).

\bibitem{Sirker}
J. Sirker, M. Maiti, N. P. Konstantinidis, and N. Sedlmayr, J. Stat. Mech. P10032 (2014).

\bibitem{Taie}
S. Taie, R. Yamazaki, S. Sugawa, and Y. Takahashi, Nat. Phys. {\bf 8}, 825 (2012).

\bibitem{SPL1}
P. Soltan-Panahi, J. Struck, P. Hauke, A. Bick, W. Plenkers, G. Meineke, C. Becker, P. Windpassinger, M. Lewenstein, and K. Sengstock, Nat. Phys. {\bf 7}, 434 (2011).

\bibitem{SPL2}
B. Yang, H. N. Dai, H. Sun, A. Reingruber, Z. S. Yuan, and J. W. Pan, Phys. Rev. A {\bf 96}, 011602(R) (2017).

\bibitem{Dutta}
O. Dutta, M. Gajda, P. Hauke, M. Lewenstein, D.-S. Luhmann, B. A. Malomed, T. Sowinski, and J. Zakrzewski, Rep. Prog. Phys. {\bf 78}, 066001 (2015).

\end{thebibliography}
\end{document}